\documentclass[reprint,amsmath,amssymb,noeprint,floatfix,superscriptaddress,longbibliography]{revtex4-2}

\usepackage[english]{babel}
\usepackage[T1]{fontenc}
\usepackage[utf8]{inputenc}
\usepackage[colorinlistoftodos, color=green!40, prependcaption]{todonotes}
\usepackage{graphicx}
\usepackage{yhmath}
\usepackage{subfigure}
\usepackage{mathdots}
\usepackage{MnSymbol}
\usepackage{dsfont}
\usepackage{soul}
\usepackage[normalem]{ulem}
\usepackage{hyperref}
\usepackage{mathtools}
\hypersetup{
    unicode=true, % non-Latin characters in Acrobat?s bookmarks
    plainpages=false,
    colorlinks=true,% false: boxed links; true: colored links
    linkcolor=blue,% color of internal links
    citecolor=blue,% color of links to bibliography
    filecolor=blue,% color of file links
    urlcolor=blue% color of external links
}

\DeclareMathOperator{\rank}{rank}
\DeclareMathOperator{\adj}{adj}
\DeclareMathOperator{\mr}{mr}
\NewDocumentCommand{\vect}{m}{\mathbf{#1}}
\NewDocumentCommand{\matr}{m}{\mathbf{#1}}
\NewDocumentCommand{\prob}{m}{\mathbb{P}{\left( #1 \right)}}

\DeclarePairedDelimiter{\ceil}{\lceil}{\rceil}

\usepackage{mathtools}

\DeclarePairedDelimiterX\braket[2]{\langle}{\rangle}{#1\,\delimsize\vert\,\mathopen{}#2}

\begin{document}
\title{Efficient Computation  Using Spatial-Photonic Ising Machines: Utilizing Low-Rank and Circulant Matrix Constraints}
\author{Richard Zhipeng Wang${}^1$, James~S.~Cummins${}^1$, Marvin Syed${}^1$, Nikita Stroev${}^2$, George Pastras${}^3$, Jason Sakellariou${}^3$, Symeon Tsintzos${}^{3,4}$, Alexis Askitopoulos${}^{3,4}$, Daniele Veraldi${}^5$, Marcello Calvanese Strinati${}^6$, Silvia Gentilini${}^7$, Davide Pierangeli${}^7$, Claudio Conti${}^5$, Natalia~G.~Berloff}
\email[correspondence address: ]{N.G.Berloff@damtp.cam.ac.uk}
\affiliation{Department of Applied Mathematics and Theoretical Physics, University of Cambridge, Wilberforce Road, Cambridge CB3 0WA, United Kingdom\\
${}^2$ Department of Physics of Complex Systems, Weizmann Institute of Science, Rehovot 76100, Israel\\
${}^3$ QUBITECH, Thessalias 8, Chalandri, GR 15231 Athens, Greece \\
${}^4$ UBITECH ltd, 95B Archiepiskopou Makariou, CY 3020 Limassol, Cyprus\\
${}^5$ Department of Physics, University Sapienza, Piazzale Aldo Moro 5, Rome 00185, Italy\\
${}^6$ Research Center Enrico Fermi, Via Panisperna 89A, 00185 Rome, Italy\\
${}^7$ Institute for Complex Systems, National Research Council (ISC-CNR), Via dei Taurini 19, 00185 Rome, Italy
}

\date{\today}

\begin{abstract}
We explore the potential of spatial-photonic Ising machines (SPIMs) to address computationally intensive Ising problems that employ low-rank and circulant coupling matrices. Our results indicate that the performance of SPIMs is critically affected by the rank and precision of the coupling matrices. By developing and assessing advanced decomposition techniques, we expand the range of problems SPIMs can solve, overcoming the limitations of traditional Mattis-type matrices. Our approach accommodates a diverse array of coupling matrices, including those with inherently low ranks, applicable to complex NP-complete problems. We explore the practical benefits of low-rank approximation in optimization tasks, particularly in financial optimization, to demonstrate the real-world applications of SPIMs. Finally, we evaluate the computational limitations imposed by SPIM hardware precision and suggest strategies to optimize the performance of these systems within these constraints.
\end{abstract}

\maketitle

\section{Introduction}
 The demand for computational power to solve large-scale optimization problems is continually increasing in fields such as synthetic biology \cite{naseriApplicationCombinatorialOptimization2020}, drug discovery \cite{nicolaouMultiobjectiveOptimizationMethods2013}, machine learning \cite{gambellaOptimizationProblemsMachine2021}, and materials science \cite{kotthoffBayesianOptimizationMaterials2021,zhangBayesianOptimizationMaterials2020}. However, many optimization problems of practical interest are NP-hard, meaning the resources required to solve them grow exponentially with problem size \cite{paschosApplicationsCombinatorialOptimization2014}. At the same time, artificial intelligence systems, including large language models with a rapidly increasing number of parameters, are leading to unsustainable growth in power consumption at data centers \cite{samsiWordsWattsBenchmarking2023}. This has spurred interest in analog physical devices that can address these computational challenges with far greater power efficiency than classical computers. Various physical platforms are being explored, including exciton-polariton condensates \cite{berloffRealizingClassicalXY2017a,lagoudakisPolaritonGraphSimulator2017,kalininNetworksNonequilibriumCondensates2018,opalaNeuromorphicComputingGinzburgLandau2019,opalaTrainingNeuralNetwork2022}, lasers \cite{hoppensteadtSynchronizationLaserOscillators2000,palObservingDissipativeTopological2017,pierangeli2019large}, and degenerate optical parametric oscillators \cite{inagakiCoherentIsingMachine2016,yamamoto_coherent_2017,honjo100000spinCoherent2021}. Many of these platforms are known as Ising machines, which aim to solve an optimization problem called the Ising problem by minimizing the Ising Hamiltonian: \begin{equation} \label{eq:ising_hamiltonian} H = -\sum_{i, j}^N {J_{ij} s_i s_j} + \sum_i^N {h_i s_i}, \end{equation} where spins $s_i =\pm 1$. Although this problem originates from a model of ferromagnetism, where the first term is the coupling term with the coupling strengths determined by matrix $\mathbf J$ and the second term represents the external magnetic field of strength $\mathbf h$, many NP problems have been mapped to it with only polynomial overhead \cite{lucas2014ising}, making it highly significant beyond its original context.

Ising machines based on spatial light modulators (SLMs) have shown their effectiveness in finding the ground state of Ising Hamiltonians, mainly due to their scalability \cite{pierangeli_scalable_2021}. However, current experimental implementations of spatial-photonic Ising machines (SPIMs) primarily use Mattis-type coupling matrices \cite{pierangeli_scalable_2021, pierangeli_adiabatic_2020, pierangeli_large-scale_2019}. The Mattis-type matrix $\mathbf{J}$ is defined as the outer product of two identical vectors:
\begin{equation}
J_{ij} = \xi_i \xi_j.
\label{xixi}
\end{equation}
This formulation results in a rank-1 matrix with $N$ degrees of freedom, where $N$ is the dimensionality of the vector $\boldsymbol{\xi}$. Typically, the coupling matrix of an Ising Hamiltonian can be any real symmetric matrix with zeros on its diagonal, encompassing up to $N(N - 1)/2$ degrees of freedom and a rank that does not exceed $N$. This restriction significantly limits the variety of Ising Hamiltonians that SPIMs can effectively realize.

Recent advancements have expanded the types of matrices that SPIMs can implement, thanks to innovations like the quadrature method \cite{sun_quadrature_2022}, the correlation function method \cite{huang_antiferromagnetic_2021}, and the linear combination method \cite{yamashita_low-rank_2023}. The quadrature method faces a specific challenge: for a given coupling matrix $J_{ij}$, it is not always clear whether and how it can be decomposed into quadrature components that, when recombined, accurately reproduce the desired matrix. In contrast, the correlation function method allows for the evaluation of matrices characterized by components of the form
\begin{equation}
J_{ij} = G(i-j) \xi_i \xi_j,
\label{gg}
\end{equation}
where $G(i-j)$ can be  arbitrary function. The Ising Hamiltonian can be effectively evaluated using the correlation function method by calculating the correlation function of measured intensity values from SPIM against a distribution function $g$, derived via the inverse Fourier transform of the function $G(i-j)$. While this method broadens the range of matrices that can be represented, it introduces a limitation: the dependency of the additional factor $G$ solely on the difference between spin indices $i-j$ restricts output to circulant matrices. Consequently, this method can only represent Ising problems with periodic geometrical properties.

The linear combination method, on the other hand, theoretically allows for the representation of arbitrary matrices. This is achieved by decomposing the required coupling matrix into a linear combination of Mattis-type matrices:
\begin{equation}
J_{ij} = \sum_{k=1}^R {\lambda_k \xi^{(k)}_i \xi^{(k)}_j}.
\label{eqn:linear_combination}
\end{equation}
Each Mattis-type matrix is then sequentially realized in the SPIM, and all outputs are electronically combined to produce the desired Ising Hamiltonian with the accurate coupling matrix. Theoretically, if rank $R = N$, any arbitrary matrix $J_{ij}$ can be represented, although the number of optical adjustments and readouts required also scales as $O(N)$. Therefore, it is crucial to investigate whether computationally challenging problems of fixed rank $R$, which does not increase with problem size, can be represented. Given that $\rank(A+B) \le \rank(A) + \rank(B)$, and $\rank(\xi_i \xi_j) = 1$, the rank of $\mathbf{J}$ is bounded by $R$. This highlights the potential of solving challenging optimization problems that can be mapped to an Ising model with a low-rank coupling matrix.

This paper identifies computationally significant problems suitable for efficient implementation on SPIM hardware or other hardware with similar coupling matrix limitations. Many NP-complete problems have already been mapped to Ising models \cite{lucas_ising_2014}; however, for effective implementation on SPIMs, these Ising models must have coupling matrices that are either low-rank or circulant. In Section \ref{sec:spim}, we discuss the energy efficiency of SPIMs that can be utilized in unconventional hardware for optimization and machine learning applications. In Section \ref{sec:inherently_low_rank_prob}, we explore problems corresponding to Ising models with inherently low-rank coupling matrices and discuss the practical limitations of such models due to the increasing precision requirements of SPIM hardware. In Section \ref{sec:low_rank_approx}, we examine the feasibility of finding \emph{approximate} solutions to computational challenges by approximating them with low-rank Ising models. Notably, a practical optimization problem in finance is well approximated by this approach, allowing for efficient SPIM implementation. The potential applicability of low-rank matrices implementing a restricted Boltzmann machine on SPIM hardware is also briefly discussed. Section \ref{sec:cnp} introduces a new variant of an NP-hard problem, the constrained number partitioning (CNP) problem, highlighting its potential to achieve computational hardness with only a slowly increasing precision requirement, enabling existing SPIM hardware to address moderately sized hard problems even with limited precision. Finally, Section \ref{sec:trans_inv_prob} addresses translationally invariant problems that can be effectively resolved using the correlation function method with SPIMs \cite{huang_antiferromagnetic_2021}.

The following section discusses SPIMs' performance metrics and inherent advantages, highlighting their potential across various computational tasks.
\section{SPIM Performance, Advantages and Generality}
\label{sec:spim}
By exploiting the properties of light, such as interference and diffraction, SPIMs and other SLM-based devices perform computations in parallel, providing significant speed advantages over electronic systems.

SPIMs use spatial light modulation to emulate Ising problems, which are fundamental to various optimization and machine learning tasks \cite{pierangeli2019large, pierangeli2020adiabatic}. They optically compute the Ising Hamiltonian from the phase-modulated image of an amplitude-modulated laser beam. Light incident on the $i$th site of the spatial light modulator with an amplitude $\xi_i$ is phase-modulated to take Ising spin states. SPIMs can efficiently process all-to-all interactions across tens of thousands of variables, with the computational time for calculating the Ising energy $H$ scaling as $\mathcal{O}(N)$ for $N$ spins \cite{ye2023photonic,prabhakar2023optimization}. However, SPIMs are optimized for Ising problems with either rank-one interaction matrices, Eq.~(\ref{xixi}),  or low-rank $R$ interaction matrices, Eq.~(\ref{eqn:linear_combination}), using multiplexing techniques \cite{yamashita2023low,ye2023photonic}. This still offers a computational advantage compared to conventional $\mathcal{O}(N^2)$ CPU operations if $R \ll N$. Despite the limitations to low-rank problems, using SPIMs and similar devices for combinatorial optimization has a significant advantage over other annealers.

The SPIM optical device comprises a single spatial light modulator, a camera, and a single-mode continuous-wave laser. The power consumption of an SLM (model Hamamatsu X15213 series) is $15 \text{W}$. The power consumption of a charge-coupled device camera (model Basler Ace 2R) is $5 \text{W}$. The power consumption of a laser (Thorlabs HeNe HNL210LB) is $10 \text{W}$. Thus, the overall SPIM power consumption is $30 \text{W}$. This can be compared with the $16 \text{kW}$ needed to run a D-WAVE system \cite{DWAVEwhitepaper}.

In terms of speed, it has been shown that the runtime for a number partitioning optimization problem with problem size $N = 16384$ is about $10$ minutes, which is comparable to the D-WAVE $5000+$ Advantage system at $N = 121$. These comparisons underscore the importance of developing optical annealers based on spatial light modulation \cite{prabhakar2023optimization}.

Although this paper focuses on SPIMs, the results apply to other alternative analog Ising machines that use similar methods of generating coupling matrices. Specifically, these machines may also employ SLM and techniques that allow the realization of low-rank interaction matrices through the direct implementation of rank-one matrices or the linear combination of multiple rank-one matrices. The problems we discuss in our paper extend their applicability to various analog computational platforms that share these experimental foundations.

Having established SPIMs' general performance and benefits, we now focus on a critical subset of problems—those characterized by inherently low-rank structures.
\section{Inherently Low Rank Problems}
\label{sec:inherently_low_rank_prob}
\subsection{Properties of Low Rank Graphs}
\label{sec:low_rank_graphs}
Given the advantages of optical annealers based on spatial light modulation, as previously discussed, it is crucial to understand the structure of Ising coupling matrices. Every Ising coupling matrix $\matr{J}$ can be associated with the (weighted) adjacency matrix of an underlying (weighted) graph. One can then define the rank of a graph by identifying it with the rank of its adjacency matrix. For graphs with identical connectivity, the unweighted version will generally have a different rank from the weighted graph.

In the case of an unweighted graph $G$ with $|G|$ vertices, it has been proven that $\rank(G) = 2$ if and only if $G = K_{p,q}$, where $K_{p,q}$ is the complete bipartite graph. A complete bipartite graph is one whose vertices can be partitioned into two subsets. Every vertex in one subset is connected to every vertex in the other, and no edge exists between vertices within the same subset. A complete tripartite graph $K_{p,q,r}$ is defined similarly, but with the vertices partitioned into three subsets, and it was also shown that $\rank(G) = 3$ is equivalent to $G = K_{p,q,r}$ \cite{gutman_nullity_2011}. Hence, any problems mapped to an \emph{unweighted} complete bipartite or tripartite graph can be efficiently implemented on SPIM.

To consider weighted graphs, a generalization of the rank of a graph is given by the \emph{minimum rank} $\mr(G)$, defined as
\begin{equation}
\mr(G) = \min_{a_{ij} \neq 0}{\left(\rank(\matr{J})\ |\ J_{ij}=a_{ij} \adj(G)_{ij} \right)},
\end{equation}
where $\adj(G)$ denotes the adjacency matrix of $G$ and minimization is over all real numbers. Essentially, the minimum rank $\mr(G)$ of a graph $G$ is the minimal rank that can be achieved by varying the connection weights in the graph while maintaining the same connectivity structure (i.e., vertices that are connected or unconnected must remain so). This value represents the minimum possible rank of the coupling matrix $\matr{J}$, where non-diagonal elements of $\matr{J}$ can be varied, but those elements that are zero must remain zero (while non-zero elements must remain non-zero). The maximum rank of a graph $G$ is always $|G|$, and it has been demonstrated that any rank between the maximum and minimum rank can be achieved \cite{fallat_minimum_2007}. For a weighted complete bipartite graph, $\mr(K_{p,q}) = 2$ \cite{fallat_minimum_2007}. Thus, any weighted graph with the structure of a complete bipartite graph has a rank between $2$ and $|G|$. This means that even if a computational problem has a complete bipartite graph structure, its adjacency matrix may still have a high rank if the edges are weighted, thus preventing efficient implementation on SPIM.

The authors of \cite{hamze_wishart_2020} proposed constructing Ising problems with tunably hard coupling matrices with exactly known rank. This family of constructed Ising problems is known as the Wishart planted ensemble, and they show a hardness peak for relatively small rank
\begin{equation}
R \approx 1.63 + 0.073N,
\end{equation}
where $N$ is the number of spins. This is not ideal as the required rank will increase linearly with the size of the problem $N$ to produce the hardest problems, but it could still serve as a benchmark for small-scale SPIM-type devices since at $N \approx 100$, the hardest problem only has rank $R \approx 8$.

\subsection{Weakly NP-Complete Problems and Hardware Precision Limitation}
\label{sec:naive_mapping_of_weakly_np_complete_problems}
The authors of \cite{yamashita_low-rank_2023} proposed a mapping from the knapsack problem with integer weights to an Ising problem with a coupling matrix $\matr{J}$ that can be represented by Eq.~(\ref{eqn:linear_combination}) with $\rank(\matr{J}) = 2$, which does not grow with the size of the problem.
The problem is defined as follows: 
Given a set of items, each having value $v_i$ and integer weight $w_i > 0$, we would like to find a subset of the items that maximize the total value of items in the subset while satisfying a constraint where the total weight of items in the subset is not greater than a given limit $W$.
The optimization version of the knapsack problem with integer weights is known to be NP-hard \cite{lucas_ising_2014}. Hence, it was argued that the linear combination method can efficiently implement the Ising formulation of NP-hard problems on SPIM.

Following a similar mapping strategy, we note that it is possible to use a single rank-1 Mattis-type matrix to represent the NP-complete number partitioning problem (NPP), which can be stated as follows: given a set of $N$ positive numbers, is there a partition of this set of numbers into two disjoint sets such that the sums of elements in each subset are equal? This can be easily mapped to the minimization problem of the Ising Hamiltonian $H_{\rm NPP}(\vect{s}) = \left(\sum_{i}^N n_i s_i\right)^2$, where $n_i$ are numbers given in the set, and $s_i \in \{-1, +1\}$ are Ising spins, which denote which subset $n_i$ is assigned to. This Ising Hamiltonian has a coupling matrix $J_{ij} = -n_i n_j$, which is of the Mattis-type, so one would expect that SPIM can readily implement any number partitioning problem without using any special rank-increasing methods mentioned in the introduction.

However, these two examples do not demonstrate that SPIM can efficiently implement computationally intractable NP-hard problems. For a number partitioning problem with $N$ integers in the range $[1, S]$, there exists an algorithm that solves the problem in time scaling like $\mathcal{O}(NS)$ \cite{garey1979computers}. This is known as a pseudo-polynomial time algorithm because if we consider the number of binary digits $L$ required to represent the largest integer $S$ in the problem, then it is given by $L = \ceil{\log_2{S}}$. The algorithm has running time scaling like $\mathcal{O}(N 2^L)$, which still grows exponentially as $L$ increases. 

Problems with such pseudo-polynomial time algorithms belong to the weakly NP-complete class, for which increasing problem size alone (in terms of $N$ for the number partitioning problem) is insufficient to make it computationally hard. These problems are only computationally intractable (i.e., having only an algorithm whose run time grows exponentially) if the number of digits used to represent the maximum input $L$ grows. If the number of digits representing the maximum input is allowed to grow, in both of the above mappings to the Ising Hamiltonian, the number of digits in the coupling matrix elements $J_{ij}$ will also have to grow. 

Hence, to simulate a weakly NP-complete problem whose solution requires exponentially growing resources on a classical computer, the precision of SPIM optical hardware will need to grow to encode larger integers represented by more binary digits $L$ involved in these problems. This is unlikely to be realized in experiments because the precision with which coupling matrices can be implemented in SPIM is a fixed number of significant digits, likely much smaller than problem input sizes of practical interest.

%%%%%%%%%%%%%%%%%%%%%%%%%%%%%%%
%Statistical mechanics analysis of the NPP
Given that NPP with limited binary precision is not NP-hard, studying the statistical properties of random NPP instances is still interesting because it may inform potential modifications or constraints to the problem that can increase its complexity. Historically, NPP was analyzed in \cite{mertens1998phase}, with subsequent extensive rigorous study in \cite{borgs2001phase}. It was found that the average hardness of a randomly generated NPP instance, where $N$ integers are uniformly randomly selected from the range $[1, 2^L]$, is controlled by a parameter $\kappa = \frac{L}{N}$. When $\kappa > \kappa_c$, they require $\mathcal{O}(2^N)$ operations to solve, but when $\kappa < \kappa_c$, average problem instances require $\mathcal{O}(N)$ operations to solve. We demonstrate a short basic derivation of the critical parameter $\kappa_c = 1$ in the limit of $N \rightarrow \infty$ below, which is the only parameter responsible for characterizing the phase of the problem, whether it is a ``hard'' phase ($\kappa > \kappa_c$) or ``easy'' phase ($\kappa < \kappa_c$).

One can introduce the signed discrepancy $D$ of the numbers, given the binary variables $s_i$ \cite{mertens_easiest_2005} as
\begin{equation}
D=\sum_{i=1}^N n_i s_i.
\label{signed_disrepancy}
\end{equation}
$D$ can be interpreted as the final distance to the origin of a random walker in one dimension who takes steps to the left $\left(s_i=-1\right)$ or to the right $\left(s_i=+1\right)$ with random stepsizes $\left(n_i\right)$. One can calculate the average number of walks that ends at $D$ as
\begin{equation}
\Omega(D)=\sum_{\left\{s_i\right\}}\left\langle\delta\left(D-\sum_{i=1}^N n_i s_i,\right)\right\rangle
\label{avg_number_of_walks}
\end{equation}
where $\langle\cdot\rangle$ denotes averaging over the random numbers $n$ and $\delta$ is the Kronecker-delta function. 
For fixed $\left\{s_i\right\}$ and large $N$, the distribution of $D$ can be treated as Gaussian with mean
\begin{equation}
\langle D\rangle=\langle n\rangle \sum_i^N s_i
\label{random_walk_mean}
\end{equation}
and variance
\begin{equation}
\begin{split}
    \left\langle D^2\right\rangle-\langle D\rangle^2 &=N\left(\left\langle n^2\right\rangle-\langle n\rangle^2\right) \\
    &= \frac{N}{12}2^{2L}\left( 1 + \mathcal{O}{\left( 2^{-L} \right)} \right)
\end{split}
\end{equation}
Hence, to the leading order in $L$, one can express the probability of the walk ending at a distance $D$ as
\begin{equation}
    p(D) = \frac{2\sqrt{3}}{2^L \sqrt{2\pi N}}\exp{\left( -\frac{6D^2}{2^{2L}N} \right).}
\label{pdf_of_D}
\end{equation}
To derive the explicit expression for the average number of walks that ends at a distance $D$ from the origin, $\Omega(D)$, we must consider that our random walk takes place on a 1-D lattice with lattice spacing $2$. In this setup, the walker's movements are restricted to even or odd numbers, depending on whether the sum $\sum n_j$ is even or odd. As a result, we obtain
\begin{equation}
\Omega(D)=2^N 2 p(D).
\label{avg_number_of_walks_final}
\end{equation}
One can get
\begin{equation}
\log _2 \Omega(0)=N\left(\kappa_c-\kappa\right),
\end{equation}
with the final expression
\begin{equation}
\kappa_c=1-\frac{\log _2 N}{2 N}-\frac{1}{2 N} \log _2\left(\frac{\pi}{24}\right).
\end{equation}
This value, denoted as \(\kappa_c\), is crucial for indicating the phase transition. 
When \(\kappa<\kappa_c\), on average, there exists an exponential number of perfect partitions where the discrepancy \(D = 0\); however, when \(\kappa>\kappa_c\), none exists. 

Another essential aspect of such a simple random walk model is the possibility of tracing the effects of finite size. For instance, even with a relatively small system size of around $N \approx 17$ units, the critical value of $\kappa_c \approx 0.9$ is due to the finite-size scaling window of the transition. These effects become more pronounced in hardware systems that operate with limited variables.

This statistical analysis aligns with our previous discussion, which suggested that to generate computationally hard NPP instances, the number of integers \(N\) and binary digits \(L\) must increase, which can be challenging in practice. Nevertheless, despite its limitations, NPP is a suitable platform for integrating additional modifications or constraints, making it adaptable for deployment on hardware with limited physical resources.

The multiprocessor scheduling problem \cite{bauke2003phase} stands out among the conventional NPP modifications. It involves distributing the workload across parallel computers to minimize overall runtime. The transition can be detailed by mapping it onto a mean-field anti-ferromagnetic Potts model. In the multiprocessor scheduling problem, we are given \(q\) identical processors and \(N\) independent tasks with running times \(a_i\) (instead of numbers \(n_i\)). The objective is to create a schedule to assign the \(N\) tasks to the \(q\) processors to minimize the longest task finishing time
\begin{equation}
T(s)=\max _\alpha\left\{A_\alpha=\sum_{i=1}^N a_i \delta(s(i)-\alpha)\right\},
\label{makespan}
\end{equation}
where \(s(i)\) denotes which processor task \(i\) is assigned to, and \(A_\alpha\) represents the total workload of processor \(\alpha\). The critical factor is whether two tasks are on the same processor. It is convenient to encode the schedules using Potts vectors \cite{wu1982potts} since this reflects the symmetry of the problem.
It was shown that this results in a different critical value for the easy-hard phase transition (that is less than $1$ for $q>1$):
\begin{equation}
\kappa_c=\frac{\log _2 q}{q-1}-\frac{1}{2 N} \log _2\left(\frac{2 \pi N}{3 q^{q /(q-1)}}\right).
\end{equation}
Such a shift is beneficial in our context because one requires smaller precision to achieve the same complexity on the hardware. 

A more straightforward approach is to consider the so-called constrained number partitioning problem (CNP), which will be discussed in more detail in Section \ref{sec:cnp}. It is a variation of the original NPP where, apart from splitting integers into groups with equal sums, we also aim to meet an additional requirement known as a cardinality constraint. This constraint ensures that the difference between the numbers of integers in one group and another equals a specific value.

The NPP is a good platform for various modifications, even though it is unsuitable for implementation on SPIM in its original form. Modifications can increase complexity within constrained resources and even combine them. For example, one could tackle the multiprocessor scheduling problem while adhering to cardinality constraints. This versatility allows for fine-tuning parameters to obtain tasks with specific complexity.

\subsection{Limitation of Low Rank Matrix Mapping}
\label{sec:limitations_of_low_rank_matrix_mapping}
Our investigation in Section \ref{sec:low_rank_graphs} suggests that low-rank graphs often exhibit highly constrained connectivity, such as complete bipartite or tripartite graphs. This is expected since a low-rank adjacency matrix represents a low-dimensional manifold with reduced degrees of freedom. Consequently, the problems they represent are not likely to be NP-hard. Section \ref{sec:naive_mapping_of_weakly_np_complete_problems} further indicates that to describe a problem which requires exponentially growing time to solve on a classical computer, it is necessary to either allow the rank or the precision of each matrix element to increase with the problem size. This evidence strongly suggests the following hypothesis may be true:
``It is not possible to find constant integers $L$ and $K$ such that there exists an Ising problem with coupling matrix $\matr{J}$ with rank $K$ and maximum input precision $L = \ceil*{\log_2{\left( \max_{i,j}{\left( J_{ij} \right)} \right)}}$ such that the number of operations required to find its ground state scales as $\mathcal{O}\left( 2^N \right)$, where $N$ is the number of spins in the Ising problem.''
Given this understanding, it is still possible to utilize SPIM to tackle NP-hard problems with the following two approaches:
\begin{enumerate}
\item Find \emph{approximate} solutions to hard problems by approximating them with a low-rank matrix and then solving the approximate problem with SPIM. This is discussed in Section \ref{sec:low_rank_approx}.
\item Identify NP problems whose precision requirement $L$ and rank requirement $K$ grow slowly as the problem size $N$ increases while maintaining their hardness. One possible candidate problem is presented in Section \ref{sec:cnp}.
\end{enumerate}

\section{Low Rank Approximation}
\label{sec:low_rank_approx}
Building on our understanding of low-rank problems, this section explores the practical application of low-rank approximations.
\subsection{Decomposition of Target Coupling Matrix}
Many strongly NP-complete problems have been mapped to Ising problems with only polynomial overhead \cite{lucas_ising_2014}. However, the resultant coupling matrices usually have no fixed structure beyond being real and symmetric, so a general method to decompose any target coupling matrix $\matr{J}$ into the form given by Eq.~\eqref{eqn:linear_combination} is required.
This can be achieved by singular value decomposition (SVD), which decomposes any matrices $\matr{J}$ into vectors $\vect{u}$ and $\vect{v}$ such that $J_{ij} = \sum_{k=1}^R {\lambda_k u^{(k)}_i v^{(k)}_j}$, where $R$ is the rank of the matrix $\matr{J}$. For any symmetric $\matr{J}$, it will lead to $\vect{u} = \vect{v}$, so SVD will produce the smallest possible set of Mattis-type matrices that represents the target matrix \cite{blum_foundations_2020}.

However, SVD gives no upper bound to the number of digits required in components of $\vect{u}$ or $\vect{v}$ to represent $\matr{J}$, so the problem presented in Section \ref{sec:limitations_of_low_rank_matrix_mapping} still exists. It is not possible to guarantee that a given coupling matrix $\matr{J}$ can be represented by vectors $\vect{u}^{(k)}$ whose precision is limited.

\subsection{How Fields Influence Rank}
\label{sec:how_fields_influence_rank}
Ising-type problems with a magnetic field (i.e., a term linear in the spins in the Hamiltonian) can be reduced to a problem without a field by adding an auxiliary spin. If the initial Hamiltonian is
\begin{equation}
    H(\vect{s}) = - \sum_{i,j=1}^N J_{ij} s_i s_j - \sum_{i=1}^N h_i s_i,
\end{equation}
then the corresponding Hamiltonian without field is
\begin{equation} \label{eq:Hh}
    H^{\vect{h}}(\vect{s}) = - \sum_{i,j=0}^N J^{\vect{h}}_{ij} s_i s_j,
\end{equation}
with $J^{\vect{h}}_{i0} = J^{\vect{h}}_{0i} = h_i$ and an additional free constant $h_0$. The auxiliary spin is fixed as $s_0=1$. 
What is the rank of this new coupling matrix with respect to the original one? Note that the coupling matrix $\matr{J}^{\vect{h}} $ in Eq.~\eqref{eq:Hh} is of the form
\begin{equation}
    \begin{pmatrix}
        h_0 & \vect{h} \\
        \vect{h}^T & \matr{J}
    \end{pmatrix}.
\end{equation}
As long as \(h_0 \neq 0\), via elementary row and column operations, one can transform
\begin{equation}
    \begin{pmatrix}
        h_0 & \vect{h} \\
        \vect{h}^T & \matr{J}
    \end{pmatrix} \to
    \begin{pmatrix}
        h_0 & 0 \\
        0 & \matr{J} - \frac{1}{h_0} \vect{h}^T \vect{h}
    \end{pmatrix} \, ,
\end{equation}
and hence 
\begin{equation}
    \rank \matr{J}^{\vect{h}} = \rank h_0 + \rank \left( \matr{J} - \frac{1}{h_0} \vect{h}^T \vect{h} \right)
    \leq \rank\matr{J} + 2.
    \label{eq:rank}
\end{equation}

\subsection{Low Rank Approximation of Coupling Matrices}
Given any Ising problem with external fields, we can convert it into another Ising problem with no external field and a coupling matrix with a rank at most two higher. We can then use SVD to decompose the resultant coupling matrix into the linear combination of Mattis-type matrices. This decomposition, in general, produces $N$ terms, where $N$ is the dimension of the coupling matrix, and the precision of each term is not bounded. Under low-rank approximation, we retain only the $K$ largest $\lambda_k$ terms in the sum produced by SVD, i.e.
\begin{equation}
    \Tilde{J}_{ij} = \sum_{k=1}^{K}{\lambda_k \xi_i^{(k)} \xi_j^{(k)}},
\end{equation}
where $K < \rank(\matr{J})$, and $\tilde{\matr{J}}$ is the low rank approximation of the exact coupling matrix $\matr{J}$. Because now only an approximate solution is required, the precision in $\xi_i^{(k)}$ can also be limited by truncating excess digits.

The low-rank approximation method was used in \cite{frieze_quick_1999} to find an approximate solution to the strongly NP-complete problem of maximum cut, in which one is required to find a partition of the node set $V$ of a graph $G(V, E)$ such that the partition maximizes the total weight of edges that cross the partition. It was shown that even with the precision in $\xi_i^{(k)}$ limited to multiples of $1 / (|V| K^2)$, the proposed algorithm can still find an approximate solution within $\mathcal{O}\left(|V|^2 / \sqrt{K} \right)$ of the maximum cut in time polynomial in $|V|$.
However, this problem remains largely unexplored in the context of Ising machines. It is unclear how the number of Mattis-type matrices $K$ and the truncation of $\vect{\xi}^{(k)}$ will impact the quality of approximate solutions found by Ising machines and to what extent these two quantities can trade off against each other to maintain the required quality of the approximate solution.

\subsection{Low-Rank Approximation of Random Coupling Matrices}

To investigate the feasibility of using a low-rank approximation to find approximate solutions to Ising problems with SPIM, a random interaction matrix for an Ising problem was generated and then decomposed into constituent Mattis-type matrices using singular value decomposition (SVD). Only matrices corresponding to the $K$ largest singular values $\lambda_k$ were retained, while the remaining matrices were discarded. Each element of the retained Mattis-type matrices was rounded to the nearest $2^{-L}$. This resulted in a rank $K$ approximate matrix with precision $L$. The approximate Ising problem was then solved through a simulation of SPIM, where a random group of spins was chosen at each step and flipped, and the new Ising energy of the system was calculated. The spin flips were accepted only if the new Ising energy decreased. As the simulation progressed, larger groups of random spins were chosen to avoid being trapped in local minima of the Ising energy landscape.

\begin{figure}[ht]
    \centering
    \includegraphics[width=\columnwidth]{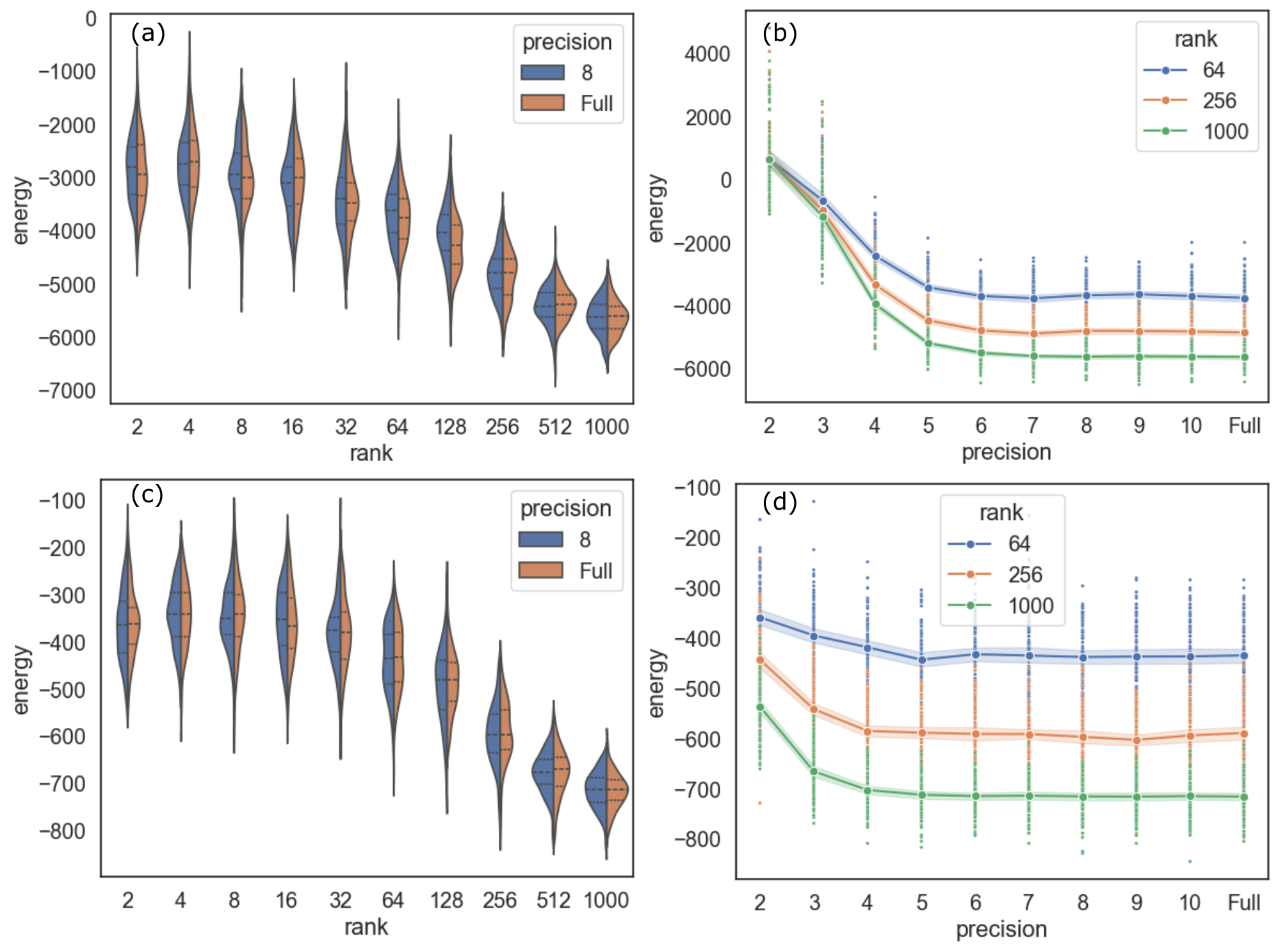}
    \caption{
(a) The energy of approximate solutions is plotted against the rank of the approximate coupling matrix. The exact interaction matrix represents a random, unweighted, undirected graph with 1000 vertices, all with anti-ferromagnetic couplings, where each edge has an equal probability of having a value of 0 or -1. Each set of parameters uses 100 different random sequences of spin flips to produce the scatter of final energies. The coupling strengths are rounded to the nearest $2^{-8}$ in the approximation. 
% {\textcolor{red}{Richard: please check if the rewording is correct.}}
(b) The energy of approximate solutions is plotted against the precision of the approximate coupling matrix. A full-rank ($R=1000$) matrix and two low-rank approximations with ranks $R=256$ and $R=64$ are used for each precision level.
(c) and (d) follow the same structure as (a) and (b), but the exact interaction matrix represents a random 3-regular graph, again with all non-zero couplings being anti-ferromagnetic.
        }
    \label{fig:low_rank_approx}
\end{figure}

Figure~\ref{fig:low_rank_approx} compares the quality of solutions obtained using varying values of $K$ and $L$. Figures~\ref{fig:low_rank_approx}(a) and (b) show results from the low-rank, limited-precision approximation of a random 1000-vertex unweighted connectivity graph, where each pair of vertices has an equal probability of being connected or unconnected. Figures~\ref{fig:low_rank_approx}(c) and (d) show results from a sparse 3-regular random graph, where each vertex is connected to 3 other random vertices. From Figs.~\ref{fig:low_rank_approx}(a) and (c), we observe that the quality of solutions obtained from 8-bit precision approximations is indistinguishable from solutions obtained from full-precision calculations, regardless of the rank of the approximate coupling matrix. This suggests that SPIM can still find highly accurate approximate solutions to Ising problems with both dense and sparse coupling matrices, even with limited precision.

This observation is further supported by Figs.~\ref{fig:low_rank_approx}(b) and (d). It can be seen that with at least 6 bits of precision, SPIM can find low-energy solutions comparable to those found by full-precision machines using the same algorithm. However, the loss of accuracy is more pronounced in dense graphs than in sparse graphs.

However, the quality of the approximate solution is highly dependent on the rank of the approximated coupling matrix. In both dense and sparse graphs, the energy of the approximate solution increases rapidly as the rank of the approximate coupling matrix decreases. Therefore, for a general random matrix, the precision of the Mattis-type matrices is not a significant factor in the quality of the approximate solution, but the rank of the approximate matrix is. The dependence on rank is likely to limit the efficiency of SPIM hardware on random Ising problems because the computation of Ising energy contributions from each rank-1 component by SPIM must be time-multiplexed, leading to much longer computation times per iteration.

In Section \ref{sec:port_opt}, we discuss a practical application in finance where a low-rank matrix can often approximate the matrix in question, making it suitable for implementation in SPIM hardware.

\subsection{Low Rank Approximation for Portfolio Optimization}
\label{sec:port_opt}

Low-rank matrices can arise during the construction of securities portfolios in financial analytics. Specifically, the optimal portfolio is the solution to a model-dependent quadratic unconstrained mixed optimization (QUMO) problem. Under an equal-weighting constraint, which we will explain below, this transforms into a quadratic unconstrained binary optimization (QUBO) problem that SPIMs can solve. Portfolio optimization involves creating an investment portfolio that balances risk and return. The objective is to allocate assets ${\bf y}_i$ optimally to maximize expected returns $\mu$ while minimizing risk $\varphi$. In modern portfolio theory, this problem is formulated in the Markowitz mean-variance optimization model \cite{markowitz1952portfolio, gilli2019numerical, beasley2013portfolio} with the objective function:

\begin{eqnarray}
H & =& (1 - \lambda) \varphi - \lambda \mu,\nonumber \\
&=& (1 - \lambda) \mathbf{w}^{T} \mathbf{S} \mathbf{w} - \lambda \mathbf{m}^{T} \mathbf{w},
\end{eqnarray}
where the scalar $\lambda \in [0, 1]$ quantifies the level of risk aversion, and $w_i \in [0, 1]$ with $\sum_{i}^{N} w_i = 1$ are portfolio weights that describe the proportion of total investment in each asset. $\mathbf{S}$ is the covariance matrix of $N$ assets with elements $S_{ij} = {\rm Cov}({\bf y}_i, {\bf y}_j)$, and $m_i$ is the expected return of asset ${\bf y}_i$. The covariance matrix $\mathbf{S}$ and return forecast vector ${\bf m}$ are derived from historical data, the latter of which can be formulated using the capital asset pricing model \cite{elbannan2015capital}. Markowitz mean-variance portfolio optimization naturally maps to the QUMO abstraction \cite{kalinin2023analog}. However, with equal weighting, the problem converts to QUBO. Equal-weighted portfolio optimization, where $w_i \in \{0, 1/q \}$ for $q$ selected assets, has been shown to outperform traditional market capitalization-weighted strategies \cite{chow2011survey, narang2013inside, swade2023equally}.

In this case, weights $w_i$ can be transformed to Ising spins $s_i \leftarrow 2q w_i - 1$. We extend the model to include a cardinality constraint that limits the portfolio to a specified number of assets, maintaining the QUBO abstraction. This is equivalent to constraining $q$ to a predetermined value. Diversification can be controlled through cardinality constraints, providing an additional mechanism to manage portfolio volatility. The objective function can be expressed as an explicit Ising Hamiltonian
\begin{equation} \label{Portfolio Equation}
\begin{split}
    H & = (1 - \lambda) \mathbf{w}^{T} \mathbf{S} \mathbf{w} - \lambda \mathbf{m}^{T} \mathbf{w} + \eta (q\mathbf{1}^{T} \mathbf{w} - q)^{2} \\
    % &= \frac{1-\lambda}{4q^2} \sum_{ij}^N {S_{ij}(s_i + 1)(s_j + 1)} \\
    % &\quad - \frac{\lambda}{2q} \sum_i^N {m_i(s_i + 1)} + \eta \left( \sum_i{(s_i + 1)} - q \right)^2 \\
    & = \sum_{i, j}^{N} {\left( \frac{1-\lambda}{4q^2}S_{ij} + 1 \right) s_i s_j} \\
    &\quad + \sum_i^N {\left( 2\eta(N-q) - \frac{\lambda m_i}{2q} + \frac{1-\lambda}{2q^2} \sum_j^N {S_{ij}} \right) s_i} + c,
\end{split}
\end{equation}
% \begin{equation}
%   \begin{alignedat}{2}
%     H  &= (1 - \lambda) \mathbf{w}^{T} \mathbf{S} \mathbf{w} - \lambda \mathbf{m}^{T} \mathbf{w} + \eta (q\mathbf{1}^{T} \mathbf{w} - q)^{2}  & &\\
%     &= bc & &= c^2
%   \end{alignedat}
% \end{equation}
%
where parameter $\eta$ controls the magnitude of the cardinality constraint and $c$ is a constant offset.
We can identify the Ising coupling matrix elements as $J_{ij} = \frac{1-\lambda}{4q^2}S_{ij} + 1$, and external magnetic field field strength as $h_i = 2\eta(N-q) - \frac{\lambda m_i}{2q} + \frac{1-\lambda}{2q^2} \sum_j^N {S_{ij}}$. By introducing an auxiliary spin to absorb linear fields with the method given in Section \ref{sec:how_fields_influence_rank} and discarding constants, the objective function becomes $H = - \mathbf{s}^{T} \mathbf{J}^{h} \mathbf{s}$. Realizing portfolio optimization in SPIM architectures requires coupling matrix $\mathbf{J}$ to be low rank. This is achieved through low-rank factor analysis (FA), a low-rank approximation technique common in quantitative finance \cite{higham2002computing}.

\begin{figure}[ht]
\centering
     \includegraphics[width=\columnwidth]{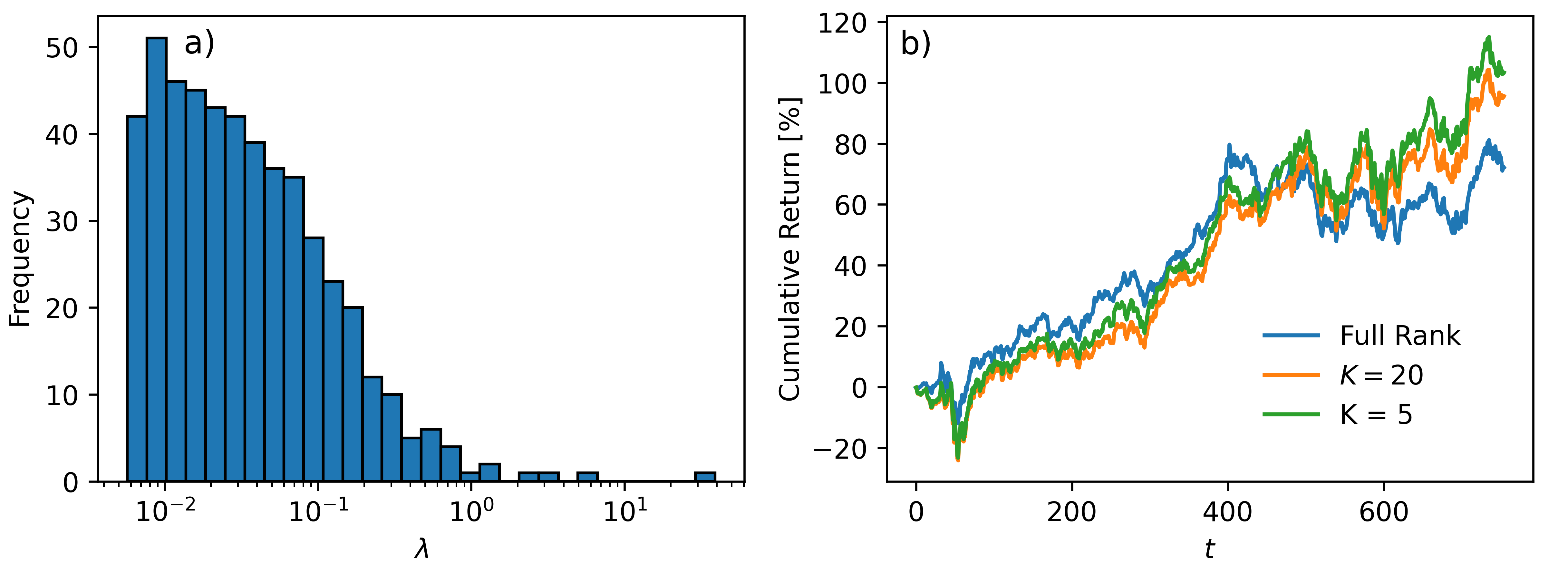}
     \caption{(a) Frequency histogram of eigenvalues obtained from the covariance matrix of S\&P 500 stock data. There are only a few dominating eigenvalues, and most eigenvalues are orders of magnitude smaller than the dominant ones. (b) Equal-weighted cardinality-constrained portfolios constructed from the full rank covariance matrix $\mathbf{S}$ (blue), $K = 20$ low rank matrix $\mathbf{S}^{'}$ (orange), and $K = 5$ low rank matrix (green). 
     Here, $\lambda = 0.5$, $\eta = 1$, and $q = 20$. The portfolios were built by minimizing Eq.~(\ref{Portfolio Equation}) using commercial solver Gurobi.}
    \label{fig:portfolio}
\end{figure}

To compute coupling matrix $\mathbf{J}$, covariance matrix $\mathbf{S}$ is first estimated from historical data on the set of asset returns $\mathbf{r}$. However, accurately estimating $\mathbf{S}$ from historical data can be challenging due to noise, high dimensionality, and limited data \cite{zhou2022covariance}. FA assumes observed data are linearly driven by a small number $K$ of common factors, such that $\mathbf{r} = \mathbf{c} + \mathbf{B} \mathbf{f} + \varepsilon$, where $\mathbf{c} \in \mathbb{R}^{N}$ is a constant vector, $\mathbf{B} \in \mathbb{R}^{N \times K}$ is the factor loading matrix with $K \ll N$, $\mathbf{f} \in \mathbb{R}^{K}$ is a vector of low-dimensional common factors and $\varepsilon \in \mathbb{R}^{N}$ is uncorrelated noise. The unobservable latent variables $\mathbf{f}$ capture the underlying patterns shared among the observed variables. FA implies the covariance matrix consists of a positive semi-definite low-rank matrix plus a diagonal matrix such that the transformed covariance matrix is $\mathbf{S}^{'} = \mathbf{B} \mathbf{B}^{T} + \Psi$ \cite{bertsimas2017certifiably}. For low-rank factor analysis, $\rank (\mathbf{B} \mathbf{B}^{T}) \leq K$ \cite{zhou2022covariance}. The diagonal matrix $\Psi$ becomes a linear field term in the binary formulation, and it follows from Eq.~(\ref{eq:rank}) that $\mathbf{S}^{'}$ remains low rank. Indeed, in the Ising QUBO abstraction with auxiliary variable, coupling matrix $\mathbf{J}^{h}$ has $\rank \mathbf{J}^{h} \leq \rank \mathbf{J} + 2 \leq \rank \mathbf{S}^{'} + \rank (\mathbf{1} \otimes \mathbf{1}^{T}) + 2 = \rank \mathbf{S}^{'} + 3$. Therefore, the transformed coupling matrix remains low rank. Figure (\ref{fig:portfolio}) illustrates the decomposition of the covariance matrix to its low-rank form and shows the proximity of portfolios constructed in the full-rank and low-rank paradigms. The full universe of stocks can be vast, so decomposing covariance matrices into low-rank forms provides computational advantages in subsequent calculations. For example, the New York stock exchange contains over 2300 stocks, whilst in Fig.~(\ref{fig:portfolio}), we consider only the 503 stocks tracked in the S\&P 500 index.

For a quadratic unconstrained continuous optimization problem, if the coupling matrix is positive semi-definite as the covariance matrix is, then the problem is convex for any linear field term. However, for QUBO problems, even if $\mathbf{S}$ and hence $\mathbf{J}$ is positive semi-definitive, the problem is not necessarily \emph{easy} to solve. The reason is that the binary constraint makes the feasible region discrete, not convex, which is why QUBO problems are generally NP-hard. SPIMs derive a temporal advantage over classical computing due to optical hardware implementing fast and energy-efficient computation. This is particularly crucial in high-frequency trading, where optimal portfolios must be calculated over microseconds to minimize latency in placing orders \cite{brogaard2014high, filipiak2017dynamic}.

\subsection{Low-Rank Matrices in Restricted Boltzmann Machines}
\label{sec:low_rank_rbm}

In Section \ref{sec:low_rank_graphs}, we mentioned that the minimum rank of a weighted complete bipartite graph is $2$. The restricted Boltzmann machine (RBM)~\cite{smolensky1986information, hinton2002training} is a computational model naturally defined on complete bipartite graphs. Thus, it is sensible to consider encoding low-rank approximations of RBMs within the SPIM paradigm.

Previous studies have shown that RBMs can provide good results when trained with low-rank approximations for collaborative filtering~\cite{salakhutdinov2007restricted}. Recent advancements have explored low-rank approximations in unrestricted Boltzmann machines (UBMs). Notably, UBMs trained on Spatial Photonic Ising Machine (SPIM) based optical hardware have demonstrated promising results. This novel approach leverages the computational power of optical systems for machine learning tasks, offering significant performance improvements~\cite{yamashita_low-rank_2023}.

Although UBMs might offer greater modeling flexibility due to their unrestricted connectivity, they suffer significant computational downsides. Each step in the UBM training algorithm involves navigating a complex energy landscape, leading to inefficient convergence. This inefficiency is particularly problematic for large-dimensional datasets, where training time and computational resources can become prohibitively high~\cite{yamashita_low-rank_2023}.

In contrast, RBMs can be framed as minimal rank problems. Their bipartite nature simplifies the training process and reduces computational complexity, making them more practical for SPIM applications. The training efficiency of RBMs was significantly enhanced by Hinton's method of minimizing contrastive divergence, which simplified the optimization of these models~\cite{hinton2002training}. This minimal rank framework suggests that RBMs can be trained more efficiently, so they are often preferred for collaborative filtering and feature learning tasks. The theoretical foundations of harmony theory, laid out by Smolensky in 1986~\cite{smolensky1986information}, underpin the information processing capabilities of both RBMs and UBMs.

In general, RBMs, exhibiting lower computational demands, are valuable tools in machine learning. Recent research demonstrating the efficacy of low-rank UBMs on SPIM hardware~\cite{yamashita_low-rank_2023} indicates that low-rank RBMs could be a promising direction for future applications of SPIM devices.

\section{Constrained Number Partitioning Problem}
\label{sec:cnp}
To further illustrate the utility of SPIMs in tackling complex problems, we introduce the constrained number partitioning problem (CNP), examining its characteristics and computational challenges.
\subsection{Definition and Characteristics of the Constrained Number Partitioning Problem}
% {\textcolor{red}{I have replaced $x_i$ with $s_i$ for consistency of spins notation and therefore bias $s$ with $S$. Please check that this notation is consistent throughout the paper.}}
In Section \ref{sec:naive_mapping_of_weakly_np_complete_problems}, it was shown that although the Number Partitioning Problem (NPP) can be mapped to an Ising problem with a rank one coupling matrix, its complexity grows as $\mathcal{O}(N2^L)$. Therefore, solving difficult NPP instances would require increasing the precision $L$ of SPIM hardware. The hardness of a random instance of NPP was characterized in \cite{walsh1996phase, mertens_easiest_2005}, which presented numerical evidence suggesting that the average complexity of a random instance is directly correlated with the probability of a perfect partition. A perfect partition is a partition where the difference in the sum of the elements in the two subsets is 0 or 1. The studies suggested that when the number of integers $N$ in the problem is large, if the probability of a perfect solution in a random instance tends to 1 (i.e., $\lim_{N \to \infty}{\prob{\text{perfect solution}}} = 1$), then the average problem is easy. Conversely, if $\lim_{N \to \infty}{\prob{\text{perfect solution}}} = 0$, then the problem is hard. It was rigorously shown that there is a phase transition separating the regimes of the asymptotic existence of a perfect solution, controlled by a parameter $\kappa = L / N$, with $\prob{\text{perfect solution}} \to 0$ when $\kappa > \kappa_c=1$. This analytical study corroborates our previous discussion that when $N$ is large, $L$ must also be large for the problem to be hard.

The authors of \cite{borgs_phase_2003} generalized these results to another problem known as the Constrained Number Partitioning (CNP) problem. In a random CNP problem, there exists a set of $N$ integers uniformly and randomly chosen in the range $[1, 2^L]$, under the constraint that the set must be partitioned into two subsets whose cardinalities differ by a given value $S$, known as the bias. The goal is to minimize the difference in the sum of elements in each subset, known as the discrepancy. Given a CNP problem with integers ${n_1, n_2, ..., n_N}$ and bias $S$, it can be mapped to an Ising problem by defining the Ising Hamiltonian
\begin{equation} \label{eq:cnp}
\begin{split}
    H &= -\left( \sum_i^N{n_i s_i} \right)^2 + A\left(\sum_i^N{s_i} - S\right)^2 \\
      &= \sum_{i,j}^N{\left(-n_i n_j + A\right) s_i s_j} - 2AS\sum_i^N{s_i} + AS^2,
\end{split}
\end{equation}
where each spin $s_i$ denotes which subset the number $n_i$ is assigned to.
The first term on the first line of Eq.~(\ref{eq:cnp}) minimizes the discrepancy between sums of two subsets, while the second term enforces the constraint that the cardinalities of subsets must differ by $S$, as long as constant $A$ is sufficiently large.
The second line of Eq.~(\ref{eq:cnp}) puts the energy into the explicit Ising form, where coupling matrix elements are $J_{ij} = n_i n_j - A$ and there is an external field with field strength $-2AS$.
By introducing an auxiliary spin as mentioned in Section \ref{sec:how_fields_influence_rank}, the external field term can be subsumed into the coupling term at the cost of increasing the rank of the coupling matrix by up to two and increasing the dimension of the coupling matrix by one.
The original coupling matrix has rank two, so this problem will have a coupling matrix of rank up to four when implemented on SPIM hardware, regardless of the number of integers in the problem.

It was found that the probability of the existence of a perfect solution for a random CNP instance is controlled by $\kappa$ and an additional parameter bias ratio $b = S/ N$.
It was rigorously shown that asymptotically as $N \to \infty$ when $b> b_c = \sqrt{2} - 1$, it is trivially easy to find the best partition because the bias is so large that it is almost always optimal to assign all largest elements to the smaller subset.
This is known as the "ordered" phase.
When $b < b_c$, the probability of existence of a perfect solution has a similar phase transition as in NPP, where $\prob{\text{perfect solution}} \to 0$ when $\kappa > \kappa_c$.
However, the critical value $\kappa_c$ was found to be a function of $b$, and $\kappa_c$ moves towards 0 as $b$ increases towards $b_c$.

This suggests that CNP can be a perfect candidate for implementation on SPIM hardware because an average random CNP instance can be computationally hard even if $L \ll N$ (i.e. $\kappa \ll 1$) as long as $b$ is sufficiently close to $b_c$.
Two areas need to be explored to establish that this problem is computationally hard and suitable for implementation on SPIM.
Firstly, the authors of \cite{borgs_phase_2003} did not rigorously show that the existence of a perfect solution is correlated with the hardness of the CNP problem instance like it is in NPP.
Secondly, in a system with finite size $N$, there will exist a non-zero value of $\kappa_{ c, {\rm min}}$ which leads to the smallest number of precision $L$ required for the average problem to be hard.
This value is obtained when bias ratio $b$ is as close as possible to $b_c$ given that $S$ must be an even or odd integer depending on $N$.
Finite-size effects are likely to make the transition between the easy and the hard phase gradual, with an intermediate region where the probability of having a perfect solution is close to neither 0 nor 1.
In the following subsection, we will numerically investigate this phase transition with a finitely sized system and understand the precision requirement for a moderately sized CNP problem that is still computationally hard.

\subsection{Computational Hardness of Random CNP Instances}

\begin{figure}[h]
    \centering
    \includegraphics[width=\columnwidth]{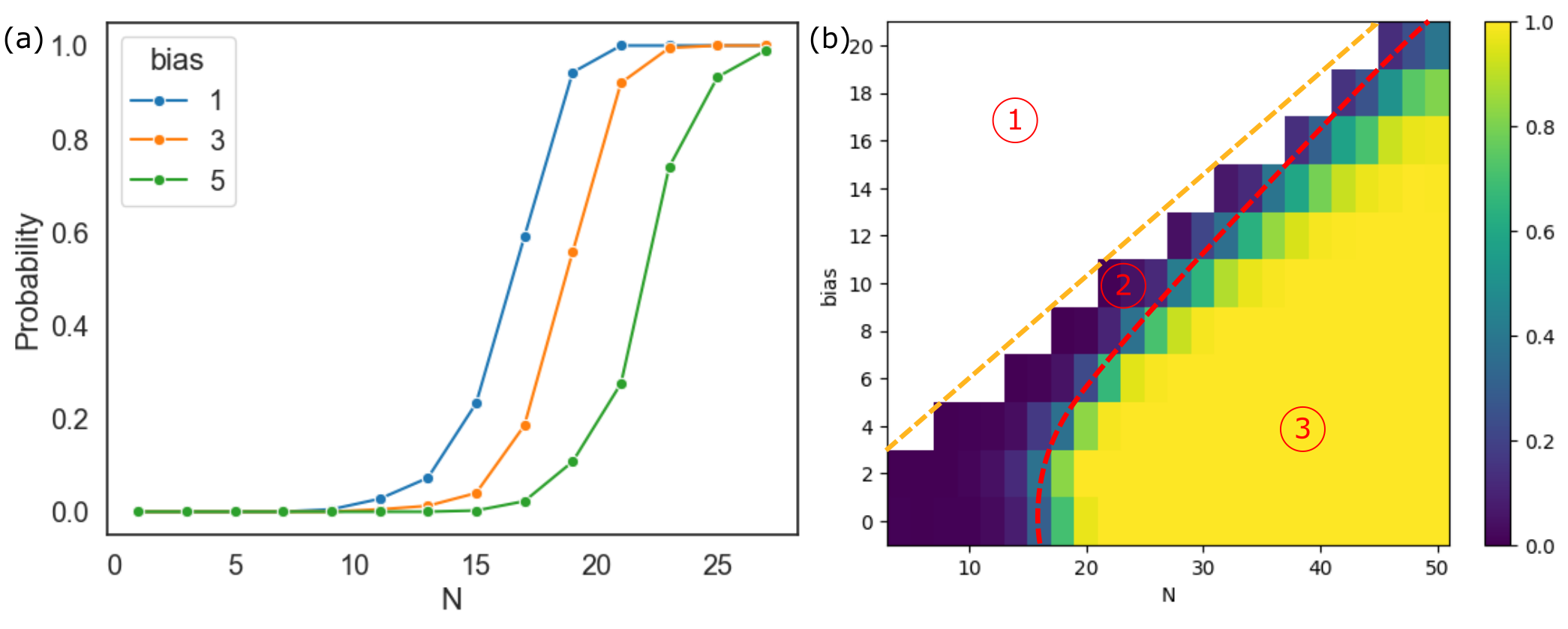}
    \caption{
        (a) The probability of the existence of a perfect solution is plotted against various problem sizes $N$ at fixed values of bias $S$.
        (b) The colour map shows the probability of the existence of a perfect solution in a random CNP problem instance with $N$ integers and various bias values, and the integers are chosen uniformly and randomly in the range $[1, 2^{12}]$.
        The probability at each point in the phase space is calculated over 200 random instances.
        Three phases are identified in the figure, separated by the orange and red dash lines. 
        Region 1, 2, and 3 correspond to the "ordered", "hard", and "perfect" phases proposed in \cite{borgs_phase_2003}.
        Region 1 in the graph is not drawn because it is likely to be trivially easy to find the optimum partition in this region for an average problem instance, so it is not meaningful to investigate the probability of a perfect solution's existence in this region.
        }
    \label{fig:perfect_solution_prob_over_bias_N}
\end{figure}

As the number of integers $N$ in a CNP problem increases, the probability of finding a perfect partition will undergo a phase transition from 0 to 1, but the critical value of $N_c$ where this happens is expected to increase as bias $S$ of the problem increases because it is increasingly challenging to balance the sums in each subset while fulfilling the constraint of greater cardinality difference.
This trend is observed in Fig.~\ref{fig:perfect_solution_prob_over_bias_N}(a).
A complete picture of the phase transition landscape is shown in Fig.~\ref{fig:perfect_solution_prob_over_bias_N}(b), which is the probability of the existence of a perfect solution in a random CNP problem with various numbers of integers $N$ and bias $S$.
We can observe a ``hard'' phase in which the probability of a perfect solution's existence is low (labelled as region 2 in the colour map) and a ``perfect'' phase in which the probability of a perfect solution's existence is close to 1 (labelled as region 3).
It can be observed that the phase transition between the ``hard'' and ``perfect'' phases is not sharp, and there exists an intermediate region where the probability of a perfect solution being present is neither close to 0 nor 1.
This can be attributed to the finite system size $N$.
Unlike the theoretical studies presented in \cite{borgs_phase_2003}, which depicts the asymptotic behaviour when $N \to \infty$, this phase diagram models a finite system implementable in SPIM hardware.
From the phase diagram, we observe that region 2, where the probability of a perfect solution's existence is close to 0, extends into $N$ values much greater than $L=12$ used in the simulation.
Hence, this numerical experiment suggests that it is possible to realize CNP problem instances with size $N$ much greater than hardware precision $L$ on SPIM and still keep the parameters in region 2, where the probability of a perfect solution being present is very low.

\begin{figure}[ht]
    \centering
    \includegraphics[width=\columnwidth]{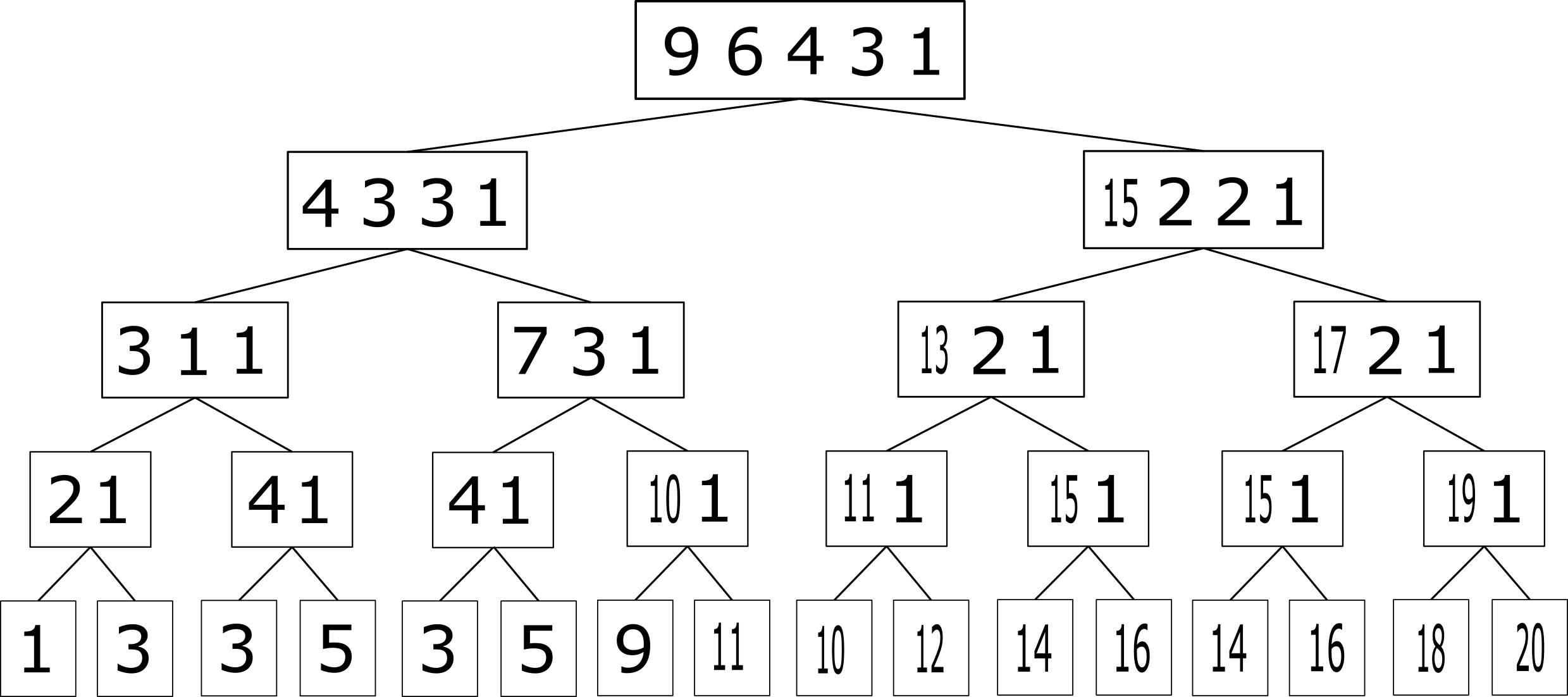}
    \caption{Schematic search tree of the complete differencing algorithm. 
    }
    \label{fig:complete_diff_algo}
\end{figure}

Next, we must understand if problem instances in region 2 represent computationally hard instances.
The existing state-of-the-art pseudo-polynomial time algorithm for number partitioning problems is the complete differencing algorithm \cite{mertens_easiest_2005}.
As shown in Fig.~\ref{fig:complete_diff_algo}, the algorithm performs the search in a depth-first manner through a tree.
The root comprises all integers in a descending order.
At each node with more than one element, the node leads to new branches.
The left branch takes the difference of the first two elements of the parent node, denoting the decision to assign the two leading elements into two different subsets.
The right branch takes the sum of the first two elements of the parent node, denoting the decision to assign the two leading elements to the same subset.
The only integer in each leaf node indicates the final discrepancy corresponding to the partition defined by the route from the root to the leaf.

This algorithm has a worst-case time complexity that grows exponentially with $N$ but relies on pruning rules to help it avoid searching through large chunks of the solution space that cannot produce a solution better than the current best-found solution.
For example, an entire branch can be discarded when the difference between the largest element and the sum of all other elements exceeds the best-known solution.
It is known that for random NPP instances that have size $N$ much greater than the precision limit $L$, this algorithm, on average, only needs to visit $\mathcal{O} (N)$ number of nodes in the search tree to find the optimal solution.
This remarkable reduction in search time from the worst case of $\mathcal{O} (2^N)$ to $\mathcal{O} (N)$ is because the probability of the existence of a perfect solution is close to 1 in the average case (when $N \gg L$). Hence, the algorithm will likely quickly find the perfect solution and terminate because no better solution is possible, thus pruning away the vast majority of the solution space.

This algorithm has also been adapted for a particular case of CNP known as the balanced number partitioning problem, where the bias $S$ is set to 0 \cite{mertens_complete_1999}.
The adapted algorithm is still efficient when $N \gg L$. This is unchanged from the NNP case - because the algorithm will likely find the perfect solution quickly and terminate before searching an exponential number of nodes in the solution space.
Hence, it is reasonable to expect that even in the case of CNP, to avoid searching the exponentially large solution space, there needs to exist exponentially many degenerate optimal solutions scattered in the solution space so that any ``good'' algorithm can quickly find one of them and terminate before visiting an exponential number of nodes.
In other words, if the number of degenerate ground states is not growing exponentially with $N$, while the total solution space is constantly increasing as $2^N$, this is a strong indication that the problem will be computationally hard.

\begin{figure}[ht]
    \centering
    \includegraphics[width=\columnwidth]{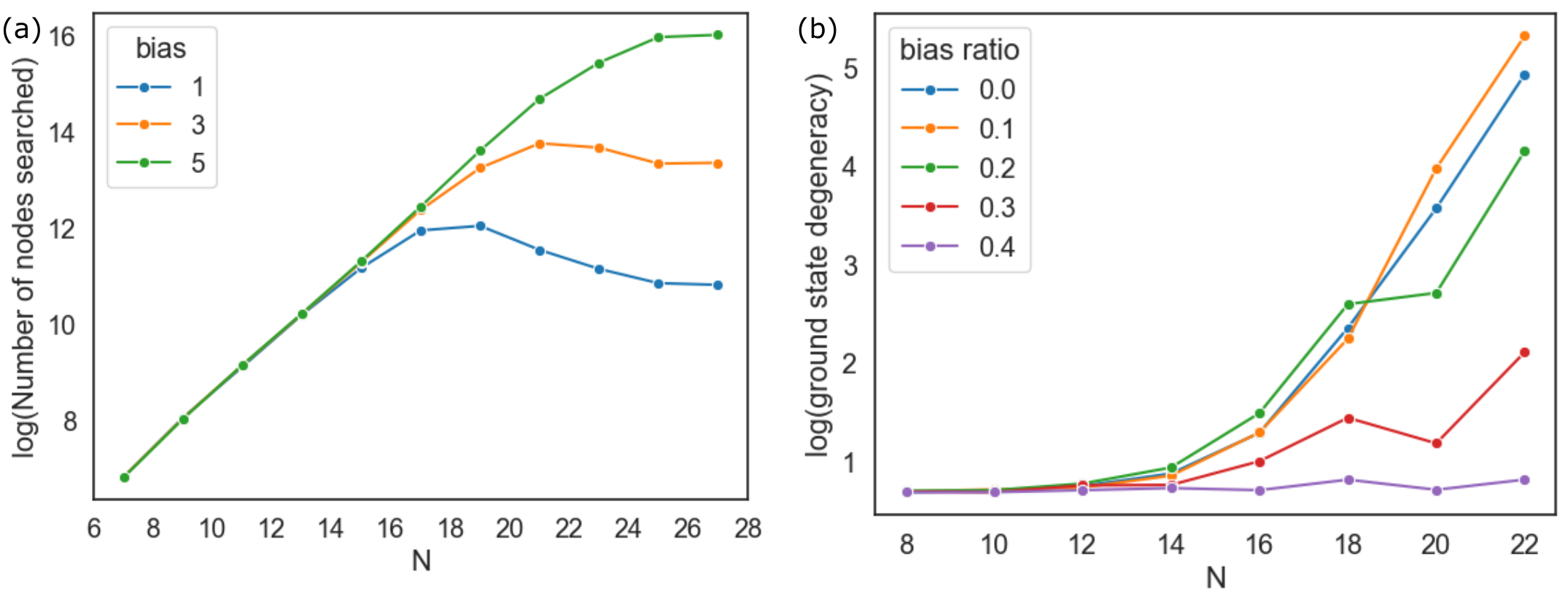}
    \caption{
        (a) The number of configurations the modified complete differencing algorithm must search to determine if a perfect solution exists in a CNP problem instance as a function of size $N$ and bias $S$.
        (b) Degeneracy of the ground state as a function of size $N$ for CNP problems with different bias ratio parameters $b$.
        Integers of the CNP instances were drawn uniformly at random from the range $[1, 2^{12}]$.
    }
    \label{fig:3_cnp_hardness}
\end{figure}

Here, we investigate two versions of the CNP problem.
The first is a decision problem: given a CNP instance, determine if a perfect solution exists.
The complete differencing algorithm was adapted by enforcing the bias constraint and then used on many random CNP problem instances with given size $N$ and bias $S$.
The number of possible configurations that must be searched before the determination can be made is shown in Fig.~\ref{fig:3_cnp_hardness}(a).
It can be observed that the number of searched configurations first increases exponentially with $N$ before hitting a peak.
We note that the average hardness of problems, as indicated by the number of configurations searched, peaks at around the same time as the phase transition from the "hard" phase to the "perfect" phase shown in Fig.~\ref{fig:perfect_solution_prob_over_bias_N}(a).
This is because if a perfect solution exists, the algorithm can locate it early on without searching the entire configuration space. If no perfect solution exists, the algorithm will likely have to search most part of the configuration space to rule it out.
Hence, our numerical test shows that CNP and NPP are likely to behave similarly, with the existence of a perfect solution correlated with the problem being easy.

The second version of CNP problems we investigate is a harder optimization problem: given a CNP instance, find the best partition, regardless of whether it is perfect.
Figure \ref{fig:3_cnp_hardness}(b) shows the degenerate ground states in randomly generated CNP problem instances with different bias parameters $b = S / N$ but with fixed precision $L = 12$.
When $N$ is smaller than $L$, it can be observed that the number of degenerate ground states did not grow exponentially with $N$ for all values of bias ratio parameters $b$, which corresponds to the lower-left corner of region 2 in Fig.~\ref{fig:perfect_solution_prob_over_bias_N}(b).
When $N$ is larger than $L$, the number of degenerate ground states grows exponentially with $N$ for smaller values of $b$.
For the largest considered $b$ value of 0.4, the number of degenerate ground states is approximately constant by an order of magnitude, even as the total solution space grows exponentially with $N$.
This strongly suggests that for the largest bias ratio $b$ value, the problem remains computationally hard even as $N$ grows while the precision $L$ is fixed.
Hence, random CNP problems with large bias ratio values can be meaningful benchmark problems for testing the performance of SPIM hardware in solving computationally hard problems because they have limited precision requirements and can be mapped to an Ising problem with a low-rank coupling matrix, as we show in the following subsection.

\section{Translation Invariant Problems}
\label{sec:trans_inv_prob}
Beyond low-rank and constrained problems, translation invariant problems offer another interesting domain for SPIM applications. This section investigates how these problems can be effectively represented and solved using SPIMs.
\subsection{``Realistic'' Spin Glass}
\begin{figure}[ht]
    \centering
    \includegraphics[width=0.3\linewidth]{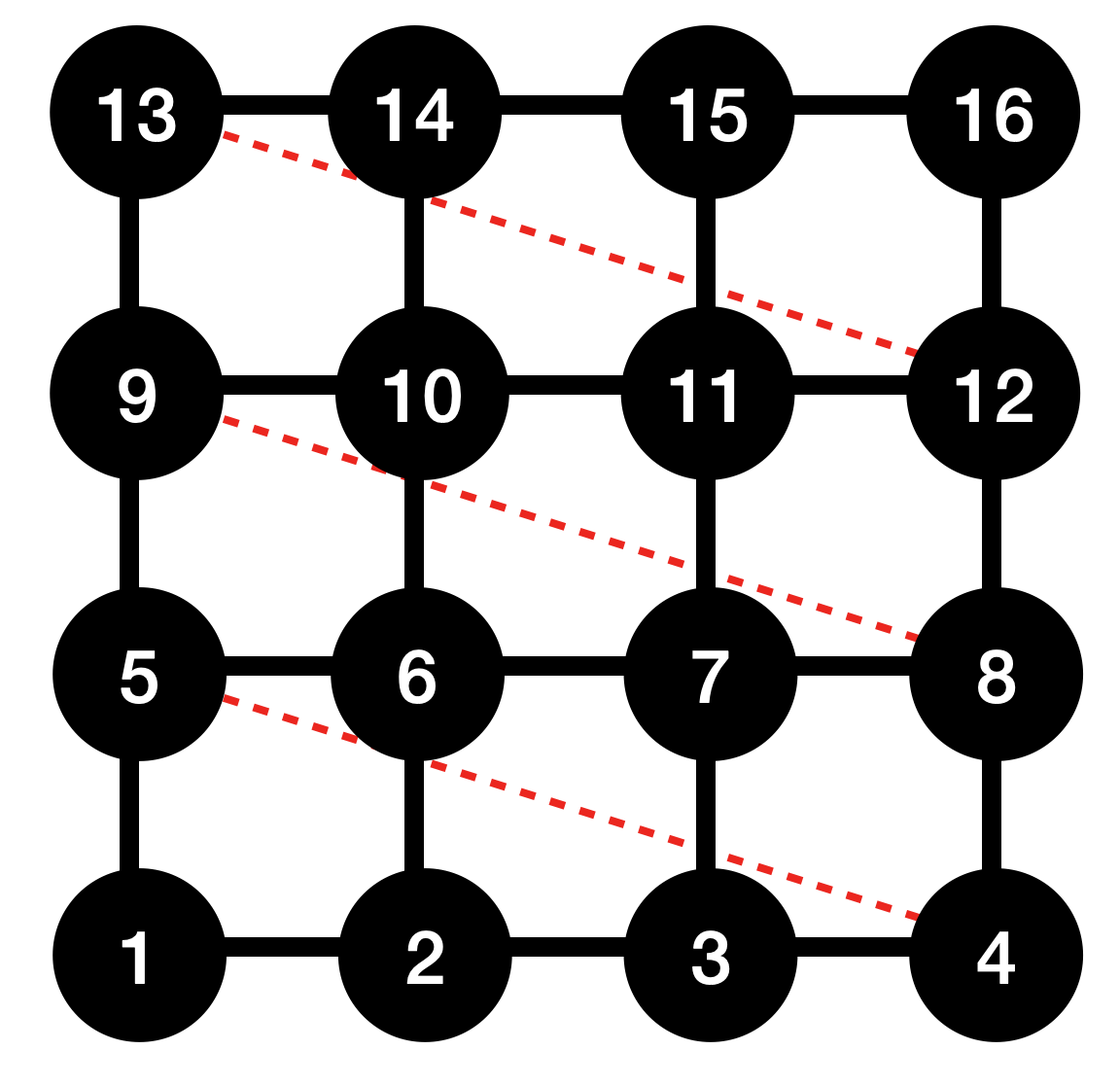}
    \caption{Connections on the \(4 \times 4\) square lattice created by the correlation function method with \(G(i-j)\) as in Eq.~\eqref{eq:realistic}. Intended connections are shown as black solid lines, while accidental connections are shown as red dashed lines.}
    \label{fig:square}
\end{figure}

The correlation function method enables SPIM to encode translation invariant (or cyclic) coupling matrices. 
This type is important, and the hard problem is ``realistic'' spin glasses that live on an almost hypercubic lattice in $d$ dimensions\,\cite{edwards1975theory, stein2013spin}. 
The modified Mattis-type matrix encoding these problems is of the form given by Eq.~(\ref{gg}),
where
\begin{equation} \label{eq:realistic}
    G(i - j) = \begin{cases}
        H_G(i - j) \text{ for } |i - j| = L^\alpha, \alpha \in \{0, 1, \dots, d-1\} \\
        0 \text{ otherwise.}
    \end{cases}
\end{equation}
Here, $H_G(k)$ could be any function. It should be noted that this encoding scheme introduces a few additional connections that do not belong to the hypercubic graph. For example, on the square $4 \times 4$ graph labelled in Fig.~\ref{fig:square}, vertices $4$ and $5$, $8$ and $9$, and $12$ and $13$ are not neighbors in this 2-dimensional lattice but would also be connected. These extra connections are thus known as ``accidental'' connections.
One could create ``glassy'' coupling matrices by choosing $H_G(k) = \cos(\omega k)$ to be sinusoidal with a suitable frequency. That is precisely how (up to decay over distance) the Ruderman-Kittel-Kasuya-Yosida exchange coupling gives rise to the anomalous magnetic behaviour measured in dilute magnetic ions in insulators that kick-started the field of spin glasses \cite{ruderman1954indirect, kasuya1956theory, yosida1957magnetic, stein2013spin}.

Another way to create frustration is to set $H_G(k) = 1$ identically but add additional anti-ferromagnetic connections to next-nearest neighbors with a similar method as in Eq.~\eqref{eq:realistic}. Spin models with such coupling are also known as \(J_1 \text{-} J_2\) models \cite{dagotto1989phase, ceccatto1992j1j2}.
Using the correlation function method, these highly frustrated coupling matrices can be realized on SPIM architecture, and the performance of sampling-based algorithms implemented on SPIM for minimizing the Ising Hamiltonian of these spin glasses can be investigated.

\subsection{Circulant Graphs}

When an $N \times N$ shift matrix $\matr{P}$  acts on a vector ${\bf x} = (x_1, x_2, \ldots, x_n)$, the components of ${\bf x}$ shift such that the order of the $x_i$ change. We describe $\matr{P}$ as \textit{cyclic} or \textit{circular} since each component $x_i$ is shifted by one around a circle. $\matr{P}^2$ turns the circle by two positions, and every new factor $\matr{P}$ gives one additional shift. $\matr{P}^N$ gives a complete $2 \pi$ shift of the components of ${\bf x}$ and therefore $\matr{P}^N = I_{N}$, where $I_{N}$ is the $N \times N$ identity matrix. A circulant matrix $\matr{C}$ is a polynomial of a shift matrix. In general
\begin{equation}
    \matr{C} = c_0 I_{N} + c_1 \matr{P} + c_2 \matr{P}^2 + \ldots + c_{N - 1} \matr{P}^{N - 1},
\end{equation}
which always has constant diagonals. The eigenvalues $\lambda$ of $\matr{P}$, given by $\matr{P} {\bf x} = \lambda {\bf x}$, are the $N$-th roots of unity. This follows from $\matr{P}^{N} = I_{N}$ to get $\lambda^{N} = 1$. The solutions are $\lambda = w, w^2, \ldots, w^{N - 1}, 1$ with $w = \exp (2 \pi i / N)$. The matrix of eigenvectors is the $N \times N$ Fourier matrix
\begin{equation}
    \matr{F} = \begin{pmatrix}
        1 & 1 & 1 & \ldots & 1\\
        1 & w & w^2 & \ldots & w^{N-1}\\
        1 & w^2 & w^4 & \ldots & w^{2(N-1)}\\
        \vdots & \vdots & \vdots & \ldots & \vdots\\
        1 & w^{N - 1} & w^{2(N-1)} & \ldots & w^{(N-1)(N-1)}
    \end{pmatrix},
\end{equation}
with orthogonal columns. Orthogonal matrices like $\matr{P}$ have orthogonal eigenvectors, and the eigenvectors of a circulant matrix are the same as the eigenvectors of the shift matrix. All information of a circulant matrix $\matr{C}$ is contained in its top row ${\bf c} = (c_0, c_1, \ldots, c_{N - 1})$, with the $N$ eigenvalues of $\matr{C}$ given by the components of the product $\matr{F} {\bf c}$. When the adjacency matrix of an undirected graph is circulant, the eigenvalues are guaranteed to be real. This is because an undirected graph has a symmetric adjacency matrix, and symmetric matrices have real eigenvalues.

\begin{figure}[ht]
    \centering
    \includegraphics[width=\columnwidth]{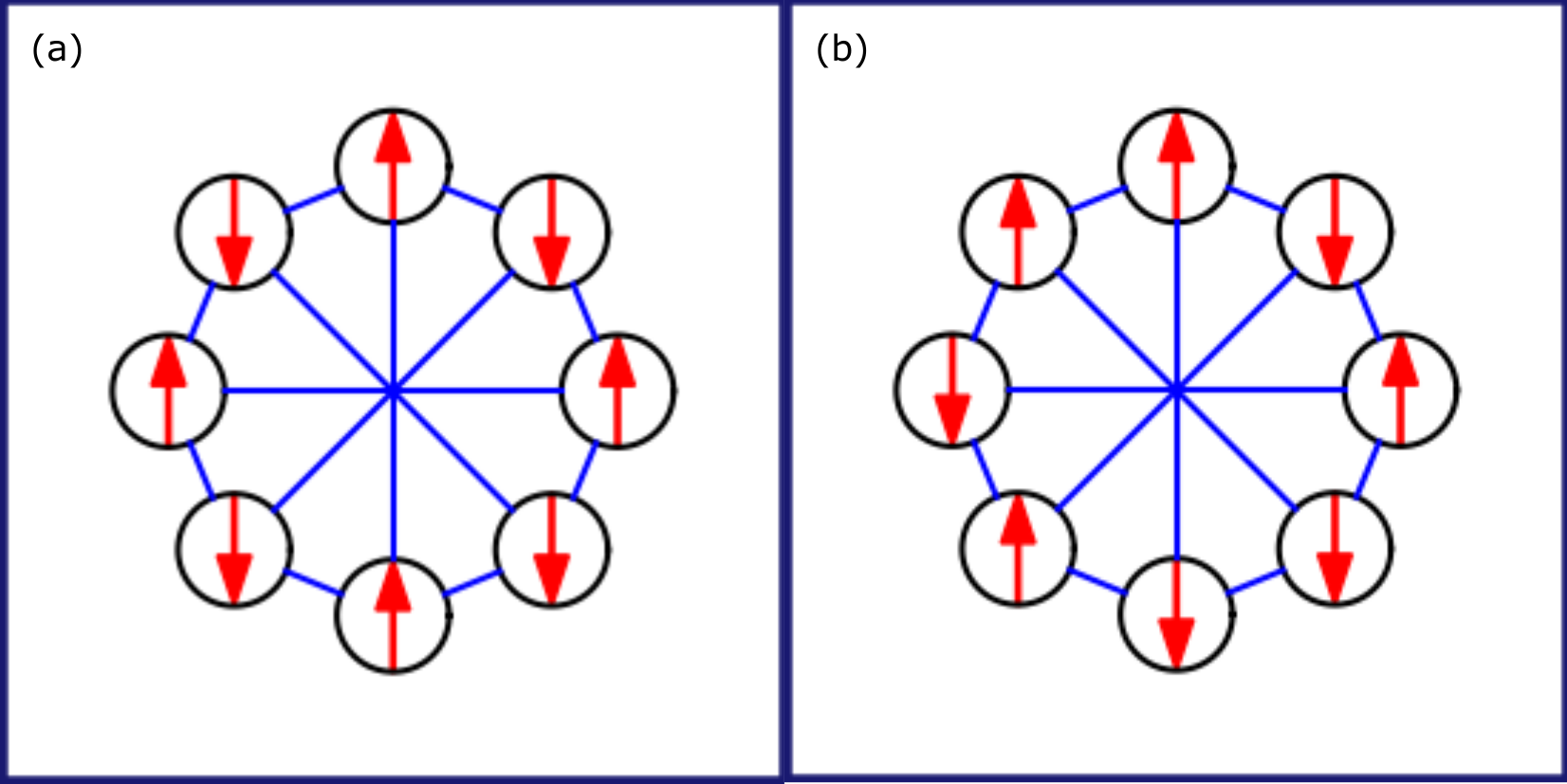}
    \caption{Possible ground state Ising spin configurations (a) State $S_0$ where each neighboring spin alternates, and (b) State $S_1$ where there are two positions, opposite on the ring, for which neighboring Ising spins are the same. For $J < J_{\rm crit}$ the neighboring couplings dominate and $S_0$ is the ground state, whereas for $J > J_{\rm crit}$ the cross-ring couplings exert a greater influence and $S_1$ is the ground state.}
    \label{fig:ising_mobius_gs}
\end{figure}

An example of a graph structure with a circulant adjacency matrix is a M\"obius ladder graph. This 3-regular graph with even number of vertices $N$ is invariant to cyclic permutations and can be implemented on SPIM hardware with each vertex of the M\"obius ladder graph representing an Ising spin. The Ising spins are coupled antiferromagnetically according to the $3N/2$ edges of the M\"obius ladder graph. Each vertex is connected to two neighboring vertices arranged in a ring, and a cross-ring connection to the vertex that is diametrically opposite, as illustrated in Fig.~(\ref{fig:ising_mobius_gs}). When $N/2$ is even, and for large cross-ring coupling, no configuration exists where all coupled Ising spins have opposite signs, and thus, frustrations must arise. The Ising Hamiltonian we seek to minimize is given by Eq.~(\ref{eq:ising_hamiltonian}) with no external magnetic field and a coupling matrix ${\bf J}$ given by the M\"obius ladder weighted adjacency matrix. The correlation function method can encode the weights of any circulant graph, which for M\"obius ladders is given by
\begin{equation}
    J_{ij} =
    \begin{cases}
    -1 &\quad \text{for } |i - j| = 1,\\
    -J &\quad \text{for } |i - j| = N/2,\\
    0 &\quad \text{otherwise}.
    \end{cases}
\end{equation}
The two types of coupling -- neighboring and cross-ring -- have different coupling strengths, the former fixed at $J_{ij} = -1$ whilst we take the latter as an adjustable parameter $J_{ij} = -J$, with $J$ constrained to the domain $[0, 1]$. The ground state takes two configurations depending on the value of $J$, as shown in Fig.~(\ref{fig:ising_mobius_gs}). The two states, denoted $S_0$ and $S_1$ have Ising energies $E_0 = (J - 2) N / 2$ and $E_1 = 4 - (J + 2) N / 2$ respectively. For the regime $J < J_{\rm crit} \equiv 4 / N$, $S_0$ is the ground state whilst for $J > J_{\rm crit}$ the energy penalty due to opposite spins having the same Ising spin sign becomes large and the ground state changes to $S_1$. For even $N/2$ and canonical shift matrix
\begin{equation}
    P = \begin{pmatrix}
        {\bf 0} & I_{N - 1}\\
        1 & {\bf 0}^{T}
    \end{pmatrix},
\end{equation}
where ${\bf 0}$ is a column vector of zeros of length $N - 1$, the weighted M\"obius ladder adjacency matrix can be expressed as ${\bf J} = - P - J P^{N / 2} - P^{N - 1}$, where the coefficients of the polynomial in $P$ are ${\bf c} = (0, -1, 0, \ldots, 0, -J, 0, \ldots, 0, -1)$. The $N$ eigenvalues of ${\bf J}$ come from multiplying the Fourier matrix $F$ with vector ${\bf c}$ to give
\begin{equation}
    \begin{pmatrix}
        \lambda_0\\
        \lambda_1\\
        \lambda_2\\
        \vdots\\
        \lambda_{N - 1}
    \end{pmatrix} = \begin{pmatrix}
        - 1 - J - 1\\
        - w - J w^{N/2} - w^{N - 1}\\
        - w^{2} - J w^{2 N / 2} - w^{2 (N - 1)}\\
        \vdots\\
        - w^{N - 1} - J w^{(N - 1) N / 2} - w^{(N - 1)(N - 1)}
    \end{pmatrix},
\end{equation}
which simplifies to $\lambda_n = - 2 \cos (2 \pi n / N) - J (-1)^{n}$. For small $J$, the first term dominates, and the largest eigenvalue is $\lambda_{N / 2} = 2 - J$. For large $J$, the second term has an effect, and the largest eigenvalues are $\lambda_{N/2 \pm 1} = 2 \cos (2 \pi / N) + J$. The eigenvectors of these eigenvalues, projected to the nearest corner of the hypercube $[-1, 1]^{N}$, correspond to states $S_0$ and $S_1$, respectively. The critical value of $J$ at which $\lambda_{N / 2} = \lambda_{N/2 \pm 1}$ occurs at $J_{\rm e} \equiv 1 - \cos(2 \pi / N)$.

Circulant graphs can be expressed as polynomials of shift matrices, from which eigenvalues and eigenvectors are calculated. This allows for a mathematically tractable analysis of the graph structure and its properties, revealing regions of parameter space for which optimization methods can falter. Moreover, circulant graphs are technologically feasible on SPIM hardware when utilizing the correlation function method \cite{huang_antiferromagnetic_2021}. To see how circulant graphs can contain non-trivial structures resistant to simple local perturbations, we note that for M\" obius ladder graphs with $J \in [J_{\rm e}, J_{\rm crit}]$ the eigenvalue corresponding to $S_0$ is less than that for $S_1$ despite $S_0$ being the lower energy (ground) state. Indeed, the authors of \cite{cummins_classical_2023} found that gradient-based soft-amplitude solvers, such as the coherent Ising machine \cite{yamamoto_coherent_2017,inagakiCoherentIsingMachine2016,honjo100000spinCoherent2021}, will encounter difficulty in recovering the ground state when $J \in [J_{\rm e}, J_{\rm crit}]$ with ground state probability decreasing as the spectral gap increases. The transformation from excited state $S_1$ to ground state $S_0$ requires $N/2$ spin flips, representing a significant energy barrier to overcome. Therefore, local perturbations are not enough to bridge the distance between hypercube corners of the ground and leading eigenvalue states. This may be overcome by using SPIM hardware paired with a sampling-based algorithm to provide feedback during each iteration of the minimization process, particularly if multiple SPIMs can be coupled to achieve a massively parallel paradigm that can efficiently sample the phase space of solutions of circulant graphs.

\section{Conclusions}

SPIMs are emerging physical computing platforms with distinct strengths and practical constraints, setting them apart from conventional digital computing technologies. As advancements in engineering and materials technology continue, these platforms are expected to see enhanced capabilities. It is, however, imperative to identify problems and methods that can effectively utilize these unique strengths, providing a robust basis for benchmarking their performance.
This paper identifies several classes of problems that are particularly well-suited for SPIM hardware. SPIMs are shown to efficiently address practical problems such as portfolio optimization through low-rank approximation techniques. This methodology also presents promising opportunities for further research, including the potential implementation of low-rank restricted Boltzmann machines on SPIMs. Furthermore, the constrained number partitioning (CNP) problem, a variation of the classic number partitioning problem, serves as a valuable benchmark for comparing the performance of SPIMs with that of classical computers. The analytically solvable circulant graph provides insights into the differences in performance between gradient-based algorithms, prevalent in many current Ising machines, and sampling-based algorithms that can be implemented on SPIMs. Additionally, SPIMs have the potential to realize many ``realistic'' spin glasses, extensively studied within the realm of statistical mechanics, thereby making numerous theoretical models experimentally viable.

Our study also highlights the importance of precision and rank in relation to the constraints of SPIM hardware. While low-rank approximations can render problems more manageable on SPIMs, the precision required for these approximations can impact computational efficiency and the accuracy of solutions. Therefore, future research must explore methods to optimize the balance between rank and precision.
Beyond portfolio optimization and Boltzmann machines, SPIMs demonstrate potential in solving various NP-hard problems through innovative mapping techniques. Advanced decomposition methods, such as singular value decomposition (SVD), enable SPIMs to manage more complex coupling matrices, expanding their applicability across different optimization tasks.

In conclusion, SPIMs represent a promising advancement in computational technologies. By focusing on low-rank approximations, constrained number partitioning, and the implementation of sophisticated algorithms, this paper sets the stage for future investigations into the capabilities and applications of SPIMs. Continued research and development in this area are crucial for fully realizing the potential of SPIMs, paving the way for novel solutions to some of the most challenging computational problems. The broader implications of this research extend to fields such as finance, logistics, and data science, where SPIMs could significantly enhance performance and efficiency, leading to substantial advancements.

\section*{Acknowledgements}

R.Z.W.\, M.S.\, G.P.\, J.S.\, A.A.\, S.T.\, C.C.\, D.P.\, S.G.\, M.C.S.\, D.V.\ and N.G.B.\ acknowledge support from HORIZON EIC-2022-PATHFINDERCHALLENGES-01 HEISINGBERG Project 101114978. 
R.Z.W.\ and N.G.B.\ acknowledge the support from the Julian Schwinger Foundation Grant No.~JSF-19-02-0005.
J.S.C.\ acknowledges the PhD support from the EPSRC. 
N.S.\ and N.G.B.\ acknowledge support from Weizmann-UK Make Connection Grant 142568. 
N.S.\ acknowledges the support provided by the Clore Fellowship. 
A.A.\ and S.T.\ acknowledge support from HoloCIM (CODEVELOP-ICT-HEALTH/0322/0047).

\bibliography{ref_new, SPIM_refs}

\end{document}